\documentclass{optica-article}

\journal{opticajournal} % for journals or Optica Open

\articletype{Research Article}

\usepackage{lineno}
\usepackage{graphicx,subcaption,float}
\usepackage{verbatim}
\usepackage{array}
\DeclareMathAlphabet      {\mathbfit}{OML}{cmm}{b}{it}

\usepackage{fancyhdr} % Přidává možnost přizpůsobení záhlaví a zápatí stránek

\begin{document}

\title{Investigation of internal electric fields in~graphene/6H-SiC under illumination by Pockels effect}

\author{Václav Dědič,\authormark{1*} Jakub Sanitrák,\authormark{1} Tomáš Fridrišek,\authormark{1} 
Martin Rejhon,\authormark{1} Bohdan Morzhuk,\authormark{1} Mykhailo Shestopalov,\authormark{1} and Jan Kunc\authormark{1}}

\address{\authormark{1}Faculty of Mathematics and Physics, Charles University, Ke Karlovu 3, 121 16, Czechia
}

\email{\authormark{*}vaclav.dedic@matfyz.cuni.cz} %% email address is required; see note below about the corresponding author designation

% use {asbstract*} to suppress the copyright line. Copyright information will be added in production

\begin{abstract*}

In this paper, we introduce a method for mapping profiles of internal electric fields in birefringent crystals based on the electro-optic Pockels effect and measuring phase differences of low-intensity polarized light. In the case of the studied 6H-SiC crystal with graphene electrodes, the experiment is significantly affected by birefringence at zero bias voltage applied to the crystal and a strong thermo-optical effect. We dealt with these phenomena by adding a Soleil-Babinet compensator and using considerations based on measurements of crystal heating under laser illumination. 
The method can be generalized and adapted to any Pockels crystal that can withstand sufficiently high voltages. We demonstrate the significant formation of space charge in semi-insulating 6H-SiC under illumination by above-bandgap light.
\end{abstract*}

%%%%%%%%%%%%%%%%%%%%%%%%%%  body  %%%%%%%%%%%%%%%%%%%%%%%%%%
%\section*{Suggested keywords (delete later)}
%silicone carbide, graphene, Pockels effect, thermo-optic effect, electric field, space charge

\section{Introduction}

Silicon carbide (SiC) is a material for the production of power semiconductor components that can work even in a harsh environment under high temperatures and/or high radiation fluxes due to its electrical, thermal, and mechanical properties~\cite{Xu2020}. Of particular note is its ability to serve as a substrate for the epitaxial growth of graphene ~\cite{Kunc2017, Kunc2018, Rejhon2019}. Focusing on the SiC photoresponse, in addition to the standard above bandgap excitation (i.e. bipolar graphene/p-SiC/n$^+$-SiC high responsivity ultraviolet phototransistor~\cite{Chava2017}, self-biased Mo/n-4H-SiC Schottky photodetectors~\cite{Chaudhuri2023} and graphene-on-SiC transistor~\cite{Waldmann2011, Waldmann2012}),
 excitation by visible and near infrared light has also been studied. Subbandgap excitation can affect the substrate conductivity through transitions and trapping at deep levels. 
Semi-insulating 4H and 6H-SiC were reported to have dominant deep levels with energies ranging between $E_c-(1.1-1.73)~\mathrm{eV}$ that affect electronic properties~\cite{Evwaraye1996, Hemmingsson1997, Mitchel1999}. These levels are related to vanadium doping and natural lattice point defects responsible for the pinning of the Fermi level. In addition to point defects, the substrate photoresponse can be affected by other crystal inhomogeneities such as the presence of stacking faults in the form of different polytypes ~\cite{Barker2018, Mandal2020}.
Lifetime-limiting defects (i.e., carbon vacancy $Z_{1/2}$ center) have also recently been reported in 4H-SiC radiation detectors~\cite{Mandal2020,Kleppinger2022}.
In ~\cite{Kelkar2008, Nunnally2010} a high power photoconductivity switch was shown based on a trap-level occupancy change related to extrinsic optical absorption in the entire volume of vanadium-doped semi-insulating 6H-SiC.
Here, under extreme conditions, the drop in SiC resistance was six orders of magnitude under Nd:YAG laser subbandgap illumination. Simulations showing related changes of the internal electric field profiles with illumination were offered in~\cite{Kelkar2008}.

The possibility of measurement of internal electric field profiles in hexagonal SiC can find applications for the optimization of power semiconductors, photonic and radiation detectors, and the study of interactions of SiC with graphene.
The internal electric field distribution reflecting the space charge in hexagonal 4H and 6H SiC polytypes can be, in principle, measured by the cross-polarizers method exploiting the linear electro-optic (Pockels) effect. Recent studies on II-VI semiconductor detectors prepared from optically isotropic cubic crystals with $\overline{4}3m$ symmetry show various applications of this electric field evaluation method related to the study of deep defect levels~\cite{Rejhon2018}, metal-semiconductor interface~\cite{Rejhon2022}, the effect of optical illumination~\cite{Washington2011,Cola2013,Cola2014} or X-rays~\cite{Prekas2010,Franc2011,Pekarek2016}, space-charge oscillations~\cite{Dedic2017}, or, nowadays, the mapping of the inhomogeneities
of the 3D electric vector field~\cite{Dedic2021,Cola2022,Cola2023}. 

However, hexagonal SiC crystals show birefringence at zero bias and the resulting non-zero transmittance through the crossed polarizers makes the standard cross-polarizers method  unusable, while it is based on monitoring of the light intensity distribution in a biased sample compared to the dark images at zero bias. 

In this paper, we show how the conventional method can be modified to 6H-SiC. Compared to the conventional method that monitors changes in light intensity, the new method monitors changes in its phase. Evaluation of the electric field is demonstrated by the measurements performed on the 6H-SiC sample equipped with graphene electrodes under dark conditions and under illumination with laser at wavelength of 405~nm. Moreover, the new method is not limited to SiC, but can be generalized and applied to other Pockels (even birefringent) crystals with sufficient resistance to higher electrical voltages. This allows to perform similar measurements on different materials and gain important insights into electric fields and interactions in a wide range of applications. The method presented here is limited to samples with rectangular prism geometry with large-area electrodes on opposite faces.

Epitaxial graphene was used as the electrode material. With the help of the method, we plan to study the possible influence of space-charge changes in SiC on the transport properties of graphene.

\section{Theory}

\subsection{Pockels effect in 6H-SiC}

The hexagonal polytypes of SiC (here 6H-SiC) belong to the $\mathrm{C}_{6v}$ crystallographic point group with $6mm$ symmetry. Without an external electric field,
their crystals are uniaxial, showing birefringence with the optical axis parallel to the $c-$axis in the (0001) direction with
ordinary $n_o=2.65$ and extraordinary $n_e=2.69$ refractive indices at 600~nm~\cite{Shaffer1971}.

The index ellipsoid of the anisotropic medium placed in the electric field $\mathbfit{E}$ is~\cite{Saleh1991} 
\begin{equation}
    \sum_{i=1}^{3}\sum_{j=1}^{3}\eta_{ij}(\mathbfit{E})x_ix_j=1,
    \label{elipsoid}
\end{equation}
in which the impermeability tensor $\eta_{ij}(\mathbfit{E})$ is
\begin{equation}
    \eta_{ij}(\mathbfit{E})= \eta_{ij}(0)+\sum_{k}r_{ijk}E_k=\eta_{ij}(0)+\sum_kr_{Ik}E_k.
\end{equation}
Here, $\eta_{ij}(0)$ is the impermeability at zero electric field, $r_{ijk}$ are the Pockels coefficients and $r_{Ik}$ are the standard reduced Pockels coefficients with indices $i,j$ reduced to $I$ due to the impermeability tensor symmetry ($\eta_{ij}=\eta_{ji}$, see more in~\cite{Saleh1991,Narasimhamurty1981}). Specifically, for the hexagonal SiC crystal with $6mm$ symmetry, it is the following:
\begin{equation}
   \eta_{ij}(0)=
   \begin{pmatrix}
   \dfrac{1}{n_o^2} & 0 & 0\\
   0 &  \dfrac{1}{n_o^2} & 0\\
   0 & 0 &  \dfrac{1}{n_e^2}
   \end{pmatrix}
   \label{eta0}
   \end{equation}
and~\cite{Narasimhamurty1981}
\begin{equation}
    r_{Ik}=
    \begin{pmatrix}
    0 & 0 & r_{13} \\
    0 & 0 & r_{13} \\
    0 & 0 & r_{33} \\
    0 & r_{42} & 0 \\ 
    r_{42} & 0 & 0 \\ 
    0 & 0 & 0
    \end{pmatrix}.
     \label{rik}
\end{equation}

\begin{figure}[t!]
\centering\includegraphics[width=0.85\textwidth]{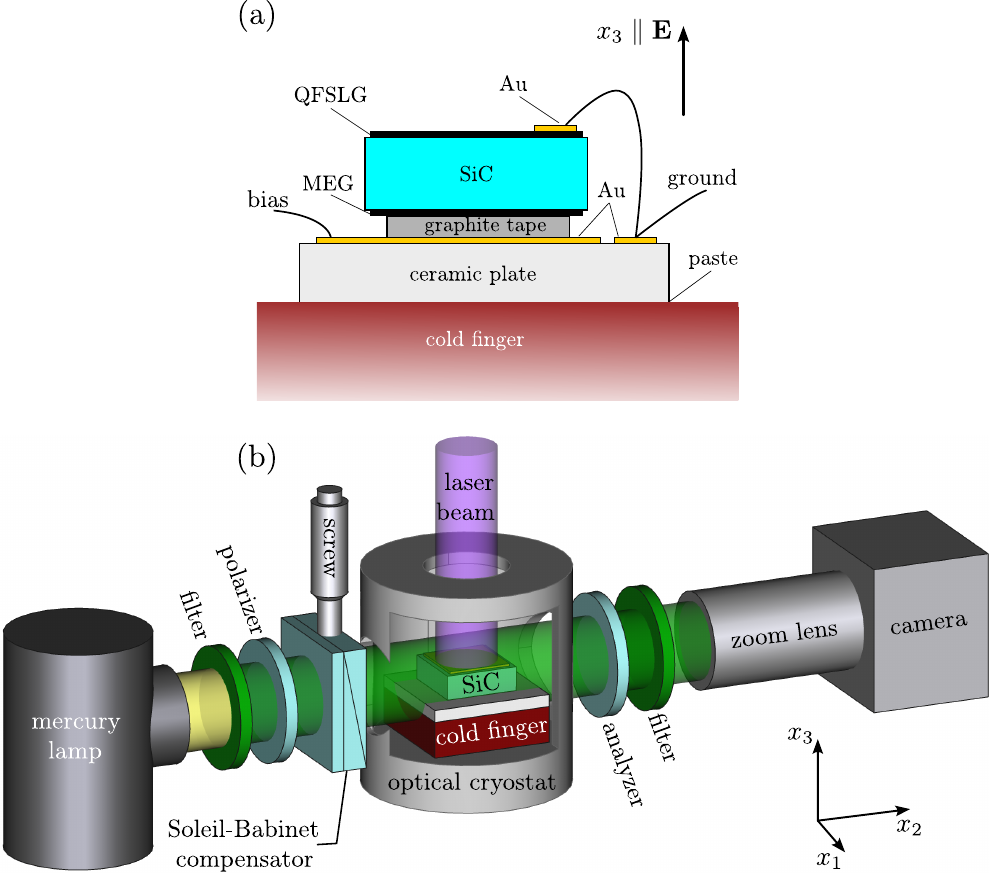}
\caption{ Scheme of the SiC sample with graphene mounted to the ceramic plate (a) with quasi free standing single layer (QFSLG) and multilayer (MEG) graphene top and bottom electrodes, respectively. (b) shows a scheme of the experimental setup used in the crossed polarizers method. 6H-SiC planar sample with graphene electrodes evaporated on large opposite sites perpendicular to the $c-$axis is placed between two orthogonal polarizers (polarizer and analyzer) oriented at 45$^\circ$ with respect to the $c-$axis. The top electrode can be illuminated by the laser. The Soleil-Babinet compensator is used for additional phase modulation. Bandpass filters at 544~nm with FWHM of 10~nm block  the scattered laser light and ambient light from hitting the CMOS camera and will simultaneously pass the low-intensity light from the 546.1~nm line of the mercury  lamp.}
\label{fig:setup}
\end{figure}

As the planar electrodes on SiC used in this study are perpendicular to the $c-$axis (and also to the $x_3-$axis, see the scheme in Fig.~\ref{fig:setup}(a)), the electric field vector is given by

\begin{equation}
\mathbfit{E}=(0,0,E).
\label{E}
\end{equation}
Combining Eqs.~\ref{elipsoid}-\ref{E}, the sought expression of the index ellipsoid is 

\begin{equation}
    \left(\dfrac{1}{n_o^2}+r_{13}E\right)\left(x_1^2+x_2^2\right)+\left(\dfrac{1}{n_e^2}+r_{33}E\right)x_3^2=1.
    \label{elipsoidsic}
\end{equation}
Eq.~\ref{elipsoidsic} can be rewritten using terms representing electric field dependent refractive indexes $1/n_o^2(E)=1/n_o^2+r_{13}E$ and $1/n_e^2(E)=1/n_e^2+r_{33}E$, which leads to a simplified form

\begin{equation}
    \dfrac{x_1^2+x_2^2}{n_o^2(E)}+\dfrac{x_3^2}{n_e^2(E)}=1,
\end{equation}
from which it is apparent that it also represents the uniaxial crystal. 
Reported experimental values of the Pockels coefficients in 6H-SiC are typically of few $\mathrm{pm/V}$~\cite{Lundquist1995,Niedermeier1999,Sato2009}.  Thus, the terms $r_{13}E$ and $r_{33}E$ from Eq.~\ref{elipsoidsic} are very low compared to $1/n_o^2$ and $1/n_e^2$, respectively, for electric field magnitudes $E$ ranging up to few kV/mm. Electric field dependent refractive indices can then be rewritten using approximation $1/\sqrt{1+\Delta}\approx 1-\frac{1}{2}\Delta$ for small $|\Delta|$~\cite{Saleh1991}  as
 \begin{equation}  
     n_e(E)\approx n_e-\dfrac{1}{2}n_e^3r_{33}E,
        \label{eq:n_e}
\end{equation}\begin{equation}
    n_o(E)\approx n_o-\dfrac{1}{2}n_o^3r_{13}E,
     \label{eq:n_o}
\end{equation}
Eq.~\ref{eq:n_e} describes the refractive index $n_e(E)$ along the $c-$axis in  direction $x_3$ and eq.~\ref{eq:n_o} describes the refractive index $n_o(E)$ in any direction perpendicular to the $c-$axis, i.e., in direction $x_1$. %including directions of $x_1$ and $x_2$ axes.

\subsection{Cross-polarizers technique}

The appropriately oriented electro-optic (EO) crystal with applied bias voltage acts as a dynamic wave retarder in which the mutual phase shift $\mathit{\Gamma}$ of two perpendicular polarizations depends on the electric field strength $E$. Generally, the electric field strength $E$ is distributed inhomogeneously between the electrodes in the studied planar samples. % (along the axis $x_3$ in Fig.~\ref{fig:setup}(a) and (b))
 On the other hand, due to the homogeneity of the crystal, the electric field remains unchanged in the directions parallel to the electrodes (axes $x_1$ and $x_2$ in Fig.~\ref{fig:setup}(b)) in the central part of the sample located far enough from the electrode edges.

Taking into account the monochromatic light (plane wave) propagating along the $x_2-$direction passing through the biased EO crystal, the mutual phase shift $\mathit{\Gamma}$ is a function of the coordinates in the $x_1x_3-$plane and the electric field amplitude $E(x_1,x_3)$ spatial distribution: \mbox{$\mathit{\Gamma}=\mathit{\Gamma}(x_1,x_3,E(x_1,x_3))$}.

This biased EO crystal placed between two perpendicular polarizers (see Fig.~\ref{fig:setup}(b)), oriented at $45^\circ$ with respect to the $c-$axis, acts as an intensity modulator with light intensity distribution~\cite{Saleh1991,Dedic2021}
\begin{equation}
    I(x_1,x_3,E(x_1,x_3))=I_0(x_1,x_3)\sin^2\dfrac{\mathit{\Gamma}(x_1,x_3,E(x_1,x_3))}{2}.
    \label{eq:T}
\end{equation}
The electric field dependent mutual phaseshift of the perpendicular polarizations at a particular point in the $x_1x_3-$plane is given by $\mathit{\Gamma}(E)=2\pi[n_o(E)-n_e(E)]d/\lambda_0$. Here $d$ is the geometric path length through the sample of the collimated low-intensity quasimonochromatic light beam from the mercury vapor lamp at the central wavelength of $\lambda_0=546.1~\mathrm{nm}$. This light source exhibits sufficient coherence to observe phenomena related to the proposed method with the chosen wavelength and sufficiently low intensity to not affect the results due to the negligible photoresponse (photocurrent$<$100~pA at 2~kV) of the studied material. 

However, during the initial experiments it turned out that an unambiguous and precise determination of the mutual phase shift $\mathit{\Gamma}(x_1,x_3,E(x_1,x_3))$ from direct intensity measurements is burdened by its periodicity and by the zero-bias birefringence. Thus, we developed a method of evaluation of the electric field by introducing a continuously variable wave retarder with known phase retardation into the optical path, namely a Soleil-Babinet (SB) compensator with the axis parallel to the $c-$axis of the EO crystal. The optical path scheme and experimental setup are shown in Fig.~\ref{fig:setup}(b).
In addition to the birefringent quartz plate, this SB compensator consists of two quartz wedges with a variable mutual geometric shift~\cite{Iizuka2002} induced by the micrometric screw with position $s$. The total thickness of the quartz wedges introduces an additional phase shift $\phi_{SB}(s)$ and the fully expressed total mutual phase shift $\mathit{\Gamma(x_1,x_3,E,s)}$ yields
\begin{equation}
\mathit{\Gamma(x_1,x_3,E,s)}=\frac{2\pi}{\lambda_0}\left[n_o-n_e+\frac{1}{2}\left(n_e^3r_{33}-n_o^3r_{13}\right)E(x_1,x_3)\right]d+\phi_{SB}(s)+aT.
 \label{eq:Gamma}
\end{equation}
Here $aT$ is temperature dependent term consisting of a thermooptic phase coefficient $a$ of a particular measured sample of EO crystal and the temperature $T$. This term is important if the temperature of the crystal varies and induces the additional phase shift which is discussed in detail in Section~\ref{thermooptic}.

\section{Experimental}
\subsection{Experimental setup}
%zde bude krome popisu aparatury i popis vzorecku

The first sample subjected to electric field measurements was a commercial (0001)-oriented semi-insulating vanadium-doped 6H-SiC with dimensions of $5\times5\times1$~mm. Both large opposite sides of the sample were equipped with semi-transparent graphene electrodes that covered the entire surface by thermal decomposition of the SiC substrate~\cite{Kunc2017, Kunc2018, Rejhon2019} in our laboratory. The top electrode was formed by a quasi free-standing single layer graphene (QFSLG) with an optical absorption of 2.3\% in the wide spectral range~\cite{Nair2008}, while the bottom electrode was formed by a multilayer graphene (MEG). The side surfaces of the sample were optically polished. The bottom sample electrode was fixed to the ceramic plate equipped with conductive channels by a conductive graphite tape, while on the top electrode there was an evaporated gold target to which a gold wire was attached by wire bonding.

\begin{figure}[h!]
\centering\includegraphics[width=0.36\textwidth]{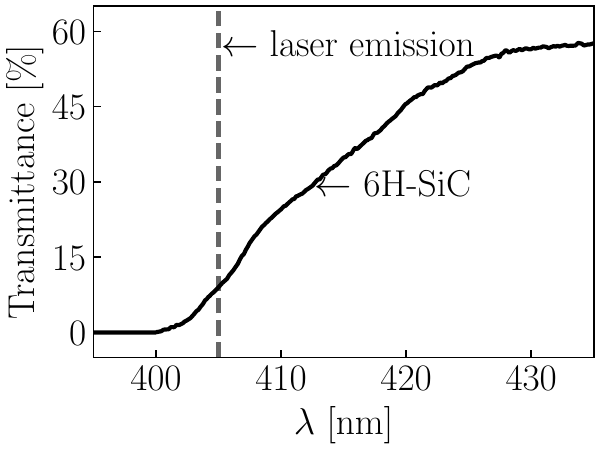}
\caption{Transmittance of the 6H-SiC wafer and the position of the 405~nm laser emission.}
\label{fig:trans}
\end{figure}

During the electric field measurements, the ceramic plate with a sample was placed on the cold finger inside the home-made optical cryostat equipped with thermoelectric temperature control located between two orthogonal polarizers (polarizer and analyzer) oriented at 45$^\circ$ with respect to the $c-$axis (see Fig.~\ref{fig:setup}(b)). Bandpass filters at 544~nm with FWHM of 10~nm block  the scattered UV and ambient light from hitting the silicon CMOS camera and will simultaneously pass the low-intensity light from the 546.1~nm line of the mercury lamp. SB compensator consists of two quartz wedges and is equipped with a motorized actuator with 1~\textmu m precision. The sample was biased by the Iseg SHQ sourcemeter and its entire top electrode was optionally illuminated by the expanded beam from the semiconductor laser Omicron at the wavelength of 405~nm. At chosen wavelength, there is enough band-to-band generation of the photocarriers in the whole sample volume, as can be observed from the spectral transmittance (see Fig.~\ref{fig:trans}) that we measured using a Fourier transform infrared spectrometer Bruker Vertex 80v on the second sample cut from the same wafer that was without graphene. We estimate the absorption of the laser light in the sample to approximately 80\% based on the wafer transmittance, expected reflectivity, and QFSLG absorption. In this study, the sample was illuminated by laser light intensities covering almost four orders of magnitude. The corresponding laser power density $S$ and the photon flux density $\Phi$, as well as estimated power $P^\mathrm{opt.}_\mathrm{absorbed}$ and number of photons absorbed by the sample, are shown in Table~\ref{tab:power}. Later in this paper, the corresponding abbreviations $S_1$ to $S_4$ are generally used for laser power densities.

The second sample without electrodes was also subjected to the measurement of the temperature increase under laser illumination. The standard platinum temperature sensor Pt100 was fixed to the top surface of the sample by a thermal paste and the ceramic plate was fixed to the aluminum table (see later~Fig.\ref{fig:termo_setup}(a)).

\begin{table}
\caption{405~nm laser power density and derived variables used in this study}
\begin{center}
\begin{tabular}{lcccccc}		
\hline\hline
&&\multicolumn{2}{c}{\small{Laser @405~nm performance:}}&&\multicolumn{2}{c}{\small{Sample absorption (estimation):}}\\
%\cline{2-5}
&&\small{Power density}& \small{Photon flux density}&& \small{Power}& \small{Photons} \\
\small{Abbreviation:}&&\small{$S$ [mW$\cdot$ cm$^{-2}$]}&\small{$\Phi$ [cm$^{-2}$s$^{-1}$]}&& \small{$P^\mathrm{opt.}_\mathrm{absorbed}$ [mW]}&\small{[s$^{-1}]$}	\\
\hline							
$S_0$ (dark) &	&	0& 0 && 0&0\\
    $S_1$  &	&	0.35&     $7.1\times10^{14}$ & &    0.07   &  $1.4\times10^{14}$\\
    $S_2$  &	&	2.8&      $5.7\times10^{15}$ & &    0.56   &  $1.1\times10^{15}$\\
    $S_3$  &	&	22&       $4.5\times10^{16}$ & &    4.4   &  $9.0\times10^{15}$\\
    $S_4$  &	&	180&      $3.7\times10^{17}$ & &    36   &  $7.4\times10^{16}$\\
\hline	\hline	
\label{tab:power}
\end{tabular}
\end{center}
\end{table}

\subsection{Measurement procedure}\label{sec:procedure}

During the electric field measurement with fixed bias and temperature, the position $s$ of the SB compensator micrometer screw changes (totally nine equidistant positions) in a range of 20~mm and the sample transmitted light intensity distribution $I(x_1,x_3,E(x_1,x_3))$ dependency (see Eqs.~\ref{eq:T}~and~\ref{eq:Gamma}) is recorded by the CMOS camera.

\begin{figure}[ht!]
\centering\includegraphics[width=0.7\textwidth]{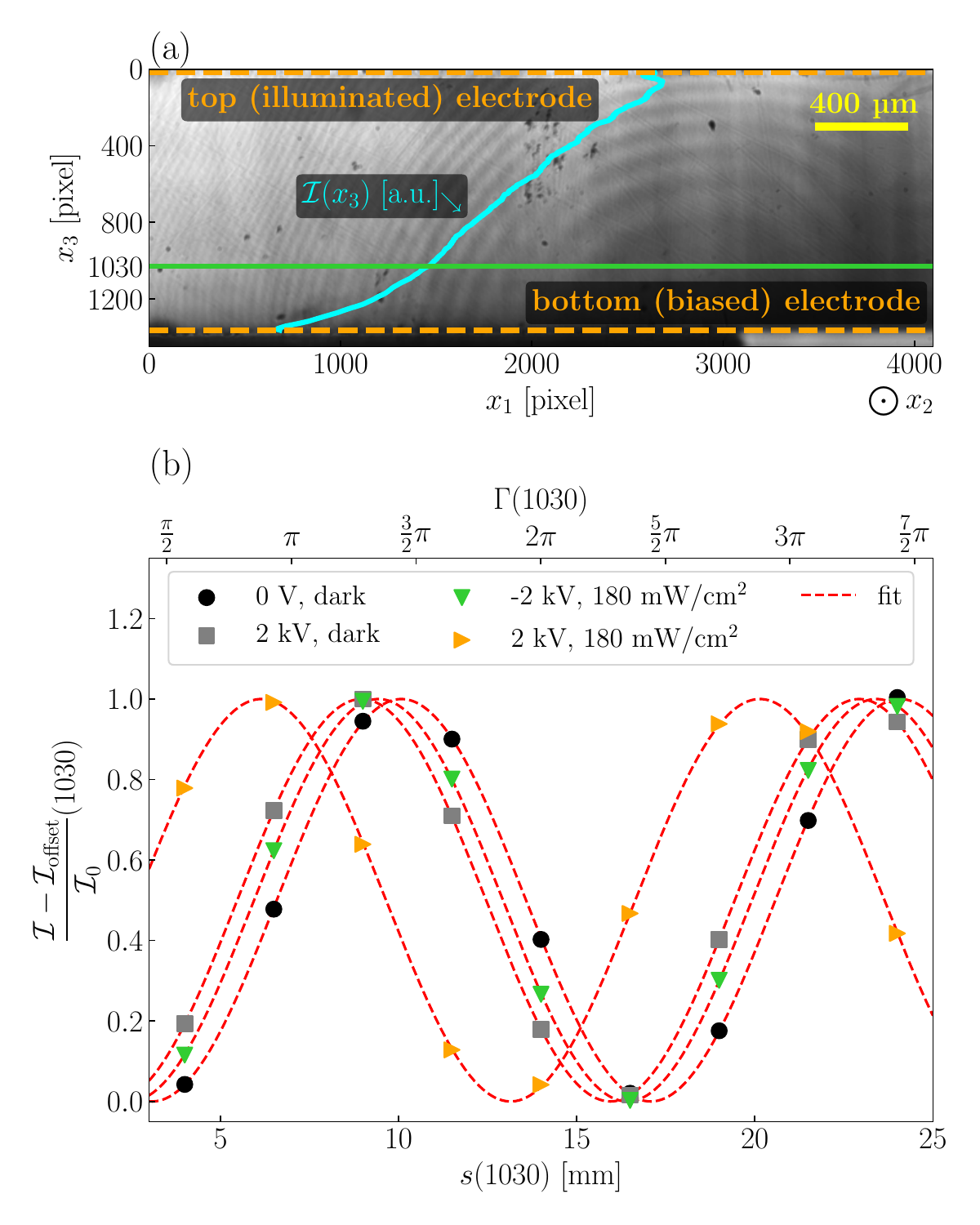}
\caption{Example of the image captured by the camera for a screw position $s=14~\mathrm{mm}$ and bias of 0~V showing the transmitted light intensity distribution $I(x_1,x_3)$ in the central part of the sample with dimensions of $\sim3.3\times 1$~mm (a). Introduction of the transmitted light intensity $\mathcal{I}(x_3)$ averaged along the $x_3-$axis (blue plot). Dark area below the bottom electrode is is caused by the light blocking by the graphite tape and the brighter part on the bottom right side is the light passing the system outside the sample volume and thus it is affected by the SB compensator only. Green line shows the pixel no.1030 in $x_3$ direction.  The data points on the plot (b) show the normalized intensity $[\mathcal{I}(1030)-\mathcal{I}_{\mathrm{offset}}(1030)]/\mathcal{I}_0(1030)$ for a pixel no.1030 on $x_3-$axis averaged  along $x_1-$axis for various screw positions $s(1030)$ of the SB compensator (bottom axis) and the corresponding phase $\Gamma(1030)$ (top axis) under different biasing and lighting conditions. The dashed curves show the fits of the experimental data according to Eq.~\ref{eq:sinusovka}.}
\label{fig:camerafit}
\end{figure}

\begin{figure}[ht!]
\centering\includegraphics[width=1\textwidth]{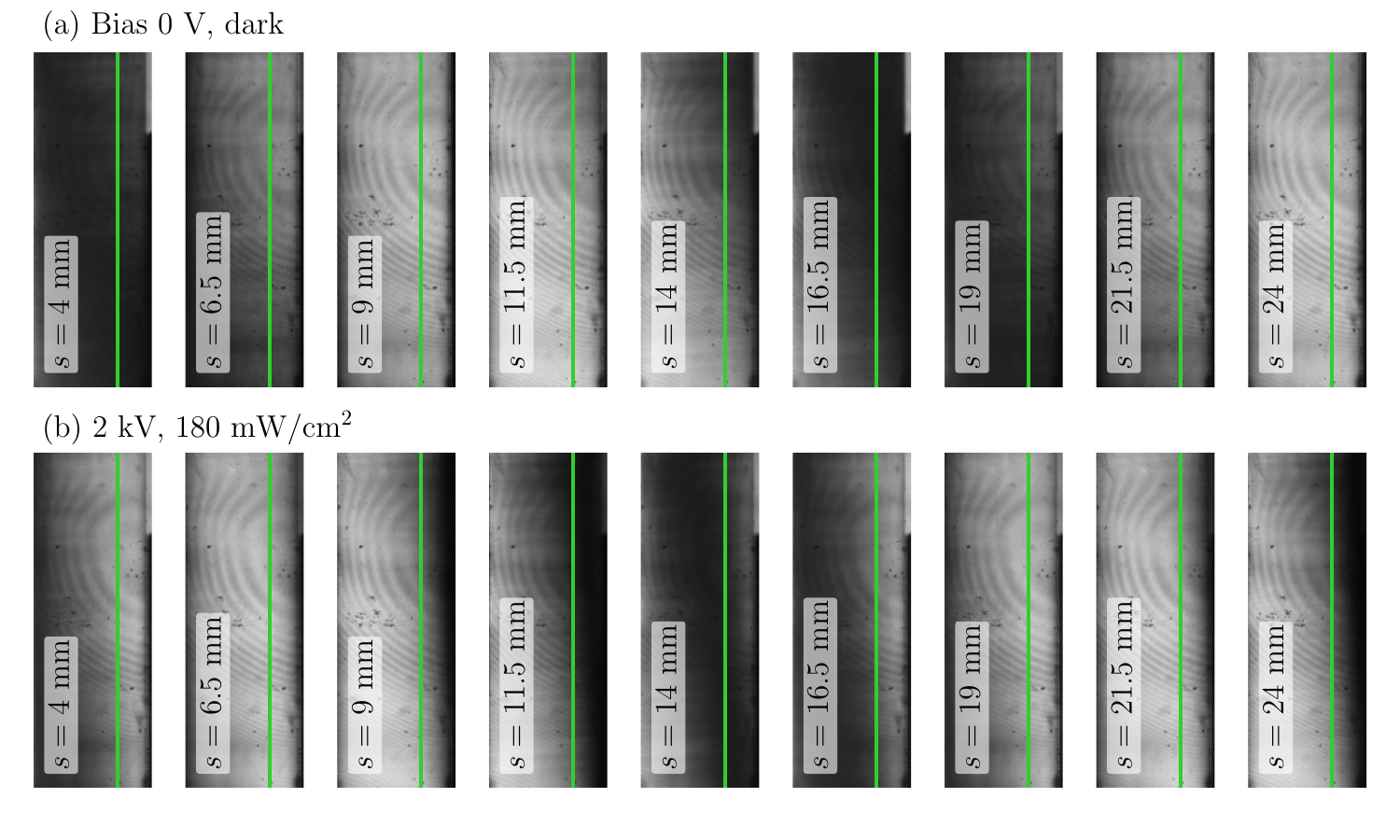}
\caption{Camera images central part of the sample, rotated by 90$^\circ$) taken for a set of screw positions $s$ of the Soleil-Babinet compensator. Dark condition at zero bias (a) and the sample biased to 2~kV with maximum 405~nm laser power density $S_4=180$~mW/cm$^2$ (b). Green lines show the pixels no.1030 in $x_3$-direction.}
\label{fig:obrazky}
\end{figure}

To simplify the evaluation of the electric field distribution, due to the geometry of the sample in the central part of the sample located far enough from the electrode edges, we expect homogeneous distributions of the electric field and the corresponding mutual phase shift along the $x_1-$direction. Therefore, new variables $\mathcal{E}(x_3)$, ${\Gamma}(x_3)$ and $\mathcal{I}(x_3)$ are introduced instead of $E(x_1,x_3)$, $\mathit{\Gamma}(x_1,x_3)$ and $I(x_1,x_3)$, respectively, with values averaged along the $x_1-$axis. An example of the image captured by the camera for a specific screw position $s$ of the SB compensator showing the transmitted light intensity distribution $I(x_1,x_3)$ in the central part of the sample with dimensions of $\sim3.3\times 1$~mm and the construction of the averaged intensity distribution $\mathcal{I}(x_3)$ is shown in Fig.~\ref{fig:camerafit}(a). The sets of camera images for various SB compensator positions $s$ demonstrating the phase shift are shown in Figs.~\ref{fig:obrazky}(a) and (b) for the dark conditions and maximum laser power, respectively.

The transmitted light intensity distribution $\mathcal{I}(x_3)$ dependency on the SB compensator position $s$ acquired by the camera is fitted at each measured point (camera pixel along the $x_3-$ axis) by the following function:
\begin{equation}
  \mathcal{I}(x_3)=\mathcal{I}_0(x_3)\sin^2\left[\dfrac{\pi}{p}(s-s_0(x_3))\right]+\mathcal{I}_{\mathrm{offset}}(x_3) 
  \label{eq:sinusovka}
\end{equation}
Here $p$ is the period corresponding to the phase $2\pi$ and $s_0(x_3)$ corresponds to the initial phase shift distribution $\Gamma_{0}(x_3)$ by 
\begin{equation}
 s_0(x_3)=\dfrac{p}{2\pi}\Gamma_{0}(x_3).\label{eq:gam0}  
\end{equation}
The fitting parameter $\mathcal{I}_{0}(x_3)$ is the amplitude. The nonzero fitting parameter $\mathcal{I}_{\mathrm{offset}}(x_3)$ is caused by imperfect light coherence, camera thermal noise, and, partially, by an inhomogeneous light distribution discussed in the last paragraph of this section. An example of the analysis described above for chosen camera pixels (average intensity $\mathcal{I}(1030)$ of pixels with no.1030 in the $x_3-$direction) is shown in Fig.~\ref{fig:camerafit}(b). We note that the coefficients of determination $R^2$ of all data fittings according to Eq.~\ref{eq:sinusovka} were greater than 0.999, indicating a relatively good precision in determining the phase shift.

From Eq.~\ref{eq:Gamma} and taking into account the constant temperature $T$, the electric field amplitude distribution $\mathcal{E}(x_3)$ is obtained from the difference of phase shift distributions $\Delta\Gamma_\mathrm{Pockels}(x_3)$ as 

\begin{equation}
  \mathcal{E}(x_3)=\dfrac{\lambda_0}{2\pi}\dfrac{\Delta\Gamma_\mathrm{Pockels}(x_3)}{\mathcal{R}d},
  \label{eq:E}
\end{equation}
where 
\begin{equation}\Delta\Gamma_\mathrm{Pockels}(x_3)=\Gamma_{0}(x_3,V)-\Gamma_{0}(x_3,0)\label{eq:subtraction}
\end{equation}is the difference of the initial  phase shift distributions in the biased ($V$) and unbiased sample, respectively. 
Term 
\begin{equation}
\mathcal{R}=\dfrac{1}{2}\left(n_e^3r_{33}-n_o^3r_{13}\right)
\end{equation}
is given by the calibration of the electric field distribution $\mathcal{E}(x_3)$ to the applied bias $V$\begin{equation}\int^t_0 \mathcal{E}(x_3)\,dx_3=V,\label{eq:calib}\end{equation}
in~which $t$ is the sample thickness.

Here we comment on significant features of the camera images shown in Figs.~\ref{fig:camerafit}(a) and~\ref{fig:obrazky}. First, there are observable scratches and several dark dots related to the surface of the sample after polishing or the dust particles in the optical setup. Second, the large concentric circular patterns with varying intensity are caused by the Newton's ring effect of the coherent light on the camera lens. Third, an inhomogeneous light intensity distribution if we compare the top left (brighter) and bottom right (darker) parts of the sample image in Fig.~\ref{fig:camerafit}(a) can originate in a combination of different phase shift by birefringence caused by variation in the geometric path $d$ caused by imperfect planparallelism of the sample and of the inhomogeneous intensity distribution of the light source incident on the camera. In this (third) case, although averaging the light intensity distribution along the $x_1$-axis can cause a reduction of the amplitude $\mathcal{I}_0(x_3)$ and an increase in the background value $\mathcal{I}_\mathrm{offset}(x_3)$, the use of the averaged $\mathcal{I}(x_3)$ instead of $I(x_1,x_3)$ still brings a precise and smooth light intensity dependence on the position of the SB compensator screw and greatly reduces the complexity of the electric field analysis.
All the features listed above are related to the particular experimental arrangement and do not affect the electric field evaluation. Their effect remains unchanged through all the electric field measurements, and it is automatically subtracted based on Eq.~\ref{eq:subtraction}.

\section{Results and discussion}

\subsection{Difference of mutual phase shift}
\begin{figure}[ht!]
\centering\includegraphics[width=1\textwidth]{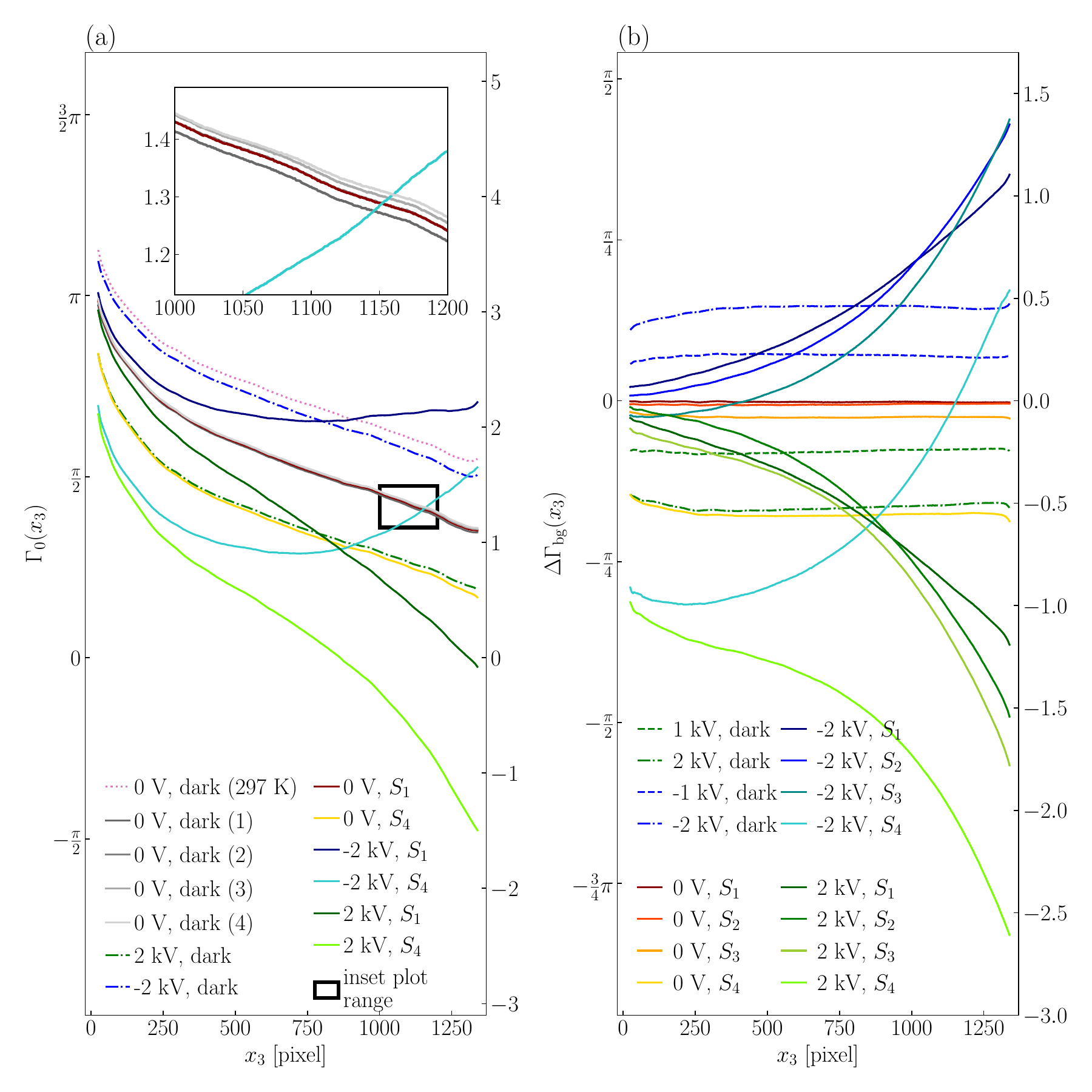}
\caption{Chosen raw mutual phase shift profiles evaluated from camera images (a). Phase shift difference profiles with a subtracted background (b). The top electrode $(x_3=0)$ was illuminated, while bias polarity is related to the bottom electrode.}
\label{bigfig1}
\end{figure}

Fig.~\ref{bigfig1}(a) shows the results of several  mutual phase shift distributions $\Gamma_0(x_3)$ evaluated using Eq.~\ref{eq:gam0} from fits (Eq.~\ref{eq:sinusovka}) of measurements performed on the sample with graphene electrodes under dark conditions and bias
voltages at $\pm2$~kV and under illumination of the top electrode ($x_3=0$) for laser 405~nm power densities $S_1$ and $S_4$ at zero bias and for $\pm2$~kV.
At the beginning, $\Gamma_0(x_3)$ was measured at an ambient laboratory temperature of 297~K, while the rest of the measurements were performed with temperature stabilization set to 300~K.
%Although this Fig.~\ref{bigfig1}(a) showing the "raw" data may seem confusing, without focusing on the details it demonstrates the basic features of the experiment.

Measurements of the background (0~V, dark) were taken four times to estimate repeatability of the experiment and the experimental error $\sigma_{\Gamma_0}$ of the phase shift determination which yields  $\sigma_{\Gamma_0}\approx0.03$~radians (see grey curves in the inset plot). Compared to these background profiles, profiles under dark condition at $\pm2$~kV are horizontally shifted depending on the voltage amplitude and the polarity. Profile at zero bias for the illumination with higher laser intensity  $S_4$ is shifted down. On the other hand, phase shift profiles at $\pm2$~kV and under illumination are both curved and shifted down. By showing the profile (0~V, dark) at 297~K (dotted pink curve) we demonstrate how a relatively small change in the temperature can have a fundamental effect on the phase of light comparable to the application of high voltage. 

The general tilt of the phase shift profiles without illumination is partly due to the imperfect planar parallelism of the side walls of the sample and thus to the different influence of the birefringence at zero bias and partly due to an inhomogeneous light intensity distribution described above in Section~\ref{sec:procedure}. 
We can remove this tilting effect by subtracting the background profiles (0~V, dark) from all other profiles, resulting in the difference in phase shift distributions $\Delta\Gamma_\mathrm{bg}(x_3)$ shown in Fig.~\ref{bigfig1}(b). Then an error of $\Delta\Gamma_\mathrm{bg}(x_3)$ is $\sigma_{\Delta\Gamma_\mathrm{bg}}=\sqrt{2\sigma_{\Gamma_0}^2}\approx0.042$~ radians due to the subtraction.

From Fig.~\ref{bigfig1}(b) it is apparent that while biasing the sample under the dark condition causes a symmetric change in $\Delta\Gamma_\mathrm{bg}(x_3)$ profiles depending on its amplitude and polarity and lower intensities $S_1$ and $S_2$ laser 405~nm illumination at $\pm2$~kV causes bending of original $\Delta\Gamma_\mathrm{bg}(x_3)$ profiles measured in darkness. On the other hand, all $\Delta\Gamma_\mathrm{bg}(x_3)$ profiles at higher laser intensities $S_3$ and $S_4$ are additionally less or more shifted to the negative values. If we consider the relationship between the phase difference related to the Pockels effect $\Delta\Gamma_\mathrm{bg}(x_3)$ and the electric field $\mathcal{E}(x_3)$ from Eq.~\ref{eq:E}, this phase shift is in violation of the calibration rule described by Eq.~\ref{eq:calib}. 
This non-standard behavior cannot be explained with the help of the Pockels effect and, as we show in detail in the following section, it is related to the thermo-optic effect. As it turns out, the studied sample is heated both by the light power of the laser and by the thermal power of the electric current flowing through the biased sample. Then the measured phase shift difference distribution $\Delta\Gamma_\mathrm{bg}(x_3)$ consists of the Pockels and thermo-optic contributions $\Delta\Gamma_\mathrm{Pockels}(x_3)$ and $\Delta\Gamma_\mathrm{thermo-opt.}(x_3)$, respectively:
\begin{equation}
\Delta\Gamma_\mathrm{bg}(x_3)=\Delta\Gamma_\mathrm{Pockels}(x_3)+\Delta\Gamma_\mathrm{thermo-opt.}(x_3).\label{eq:gamy}
\end{equation}
In the following section~\ref{thermooptic} we show how these two components can be distinguished from each other and how to unambiguously determine the distribution of the electric field $\mathcal{E}(x_3)$ in biased 6H-SiC crystal.

\subsection{Thermo-optic effect vs. Pockels effect and phase corrections}\label{thermooptic}

To investigate the temperature effect on phase shift distributions induced by 405~nm laser illumination and distinguish it from the Pockels effect, we performed two basic experiments. Namely measurement of phase shift-temperature dependency and measurement of sample warming up under laser illumination.

\subsubsection{Thermo-optic effect}
\begin{figure}[ht!]
\centering\includegraphics[width=1\textwidth]{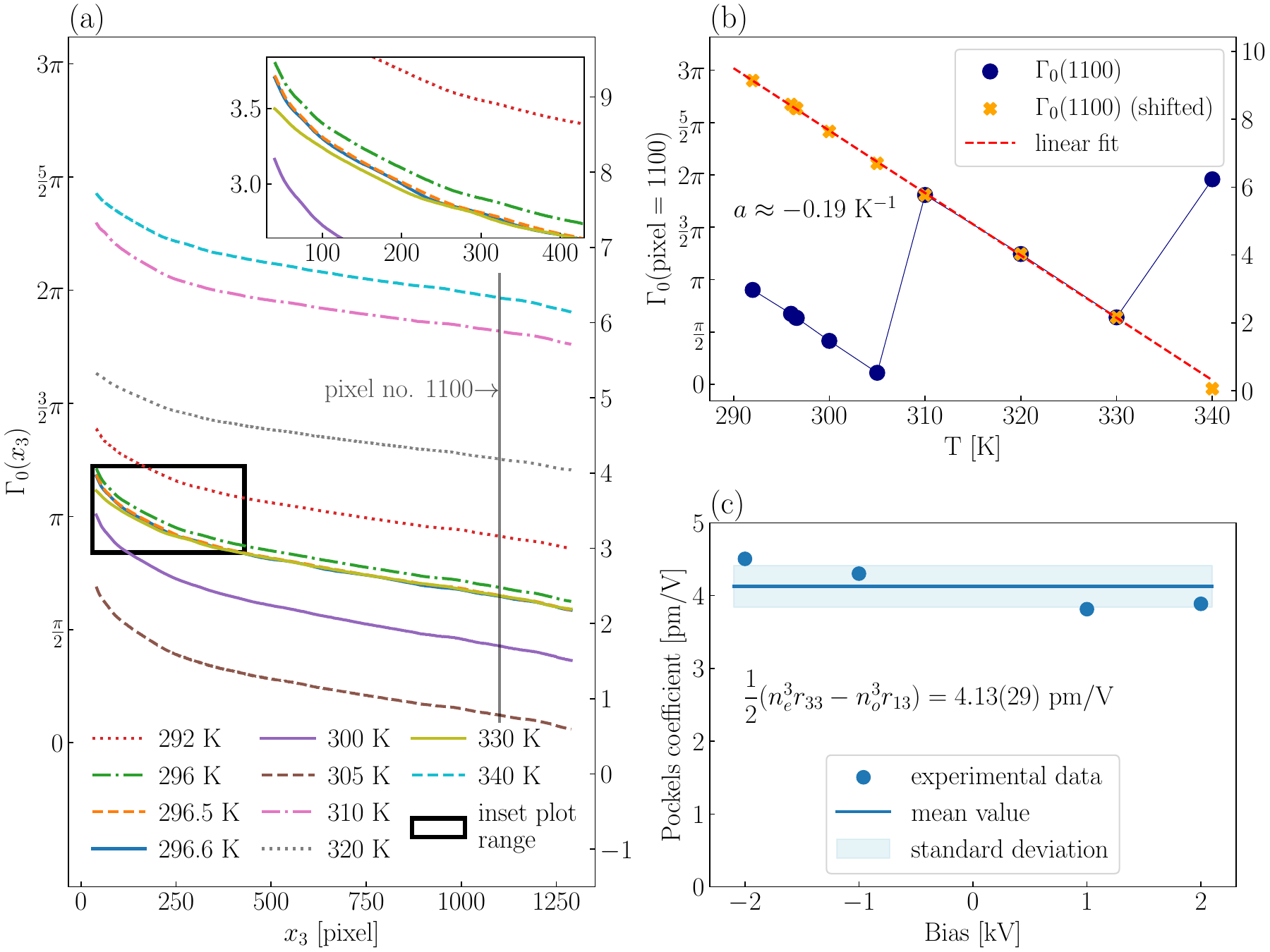}
\caption{Tepmerature dependency of $\Gamma_0(x_3)$ in unbiased sample under dark condition (a). Linear fit of manually shifted phase $\Gamma_0$ at pixel no.1100 (b). Pockels coefficient evaluated from phase shifts measured under dark conditions and under bias (c).}
\label{fig:tempdep}
\end{figure}

Fig.~\ref{fig:tempdep}(a) shows the steady state mutual phase shift distributions $\Gamma_0(x_3)$ measured at various temperature points covering range between 292~K and 340~K.  It is apparent that vertical shifts of $\Gamma_0(x_3)$ cover the whole $2\pi$ period and individual $\Gamma_0(x_3)$ profiles are almost parallel, except in the upper part of the crystal (see details in inset plot).  
We attribute this feature to the almost homogeneous temperature distribution within the sample, except for a small variation in the temperature gradient near its top surface due to the surrounding environment. From the inset plot, it is also apparent that even a small temperature drop of 0.5~K causes a significant phase difference of 0.1~radians.

The phasehift at pixel no.1100 for the whole temperature range is shown as blue circles in Fig.~\ref{fig:tempdep}(b). Its periodicity is apparent. To evaluate the temperature dependency, we shift several measured points by $\pm2\pi$ to form a straight line (orange crosses in Fig.~\ref{fig:tempdep}(b)). A slope from the linear fit of the shifted experimental data (except point at 340~K) yields the thermo-optical phase shift coefficient $a=d\Gamma_0/dT\approx-0.19~\mathrm{K}^{-1}$ for this particular crystal sample. The value of $a$ is in relatively good agreement (within 30\%) with the value calculated based on the thermo-optical coefficients of 6H-SiC published in~\cite{Xu2014}. The coefficient $a$ can be considered constant within the entire sample, due to an almost perfect parallelism of  $\Gamma_0(x_3)$ profiles from Fig.~\ref{fig:tempdep}(a). On the other hand, the roughly subtracted phase change $d\Gamma_0$ with voltage $dV$ from the Fig.~\ref{bigfig1}(b) yields $d\Gamma_0/dV\approx-0.5/2~\mathrm{kV}=-0.25~\mathrm{kV^{-1}}$, which means that the Pockels effect at the "kV per mm" scale is comparable to the thermo-optic effect in 6H-SiC for a change in several kelvin units. Therefore, stabilization of the sample temperature is essential if we take into account that the data acquisition necessary for a single mutual phase profile $\Gamma_0(x_3)$ took approximately 10 minutes and the ambient laboratory temperature can vary slightly. It should also be noted here that the measurement to obtain all the $\Delta\Gamma_\mathrm{bg}(x_3)$ profiles from Fig.~\ref{bigfig1}(b) with breaks for sample stabilization took almost 8 hours, during which the ambient temperature could change by several degrees. The optimum variation in the sample temperature during all measurements should be smaller than 0.1~K.

%nasleduje pasaz o urceni pockelsovskych koeficientu

\subsubsection{Pockels coefficients}\label{sec:R}

Using Eq.~\ref{eq:E} and the condition from Eq.~\ref{eq:calib} we evaluated the Pockels coefficient $\mathcal{R}$ for the phase-shift measurements at $\pm1$~ kV and $\pm2$~ kV under dark conditions. In the case of the measurements in the dark conditions we do not expect any significant contribution of the thermo-optic effect. The coefficient $\mathcal{R}$ should be in principle constant for 6H-SiC and the chosen light wavelength 546.1~nm. Found mean value of $\mathcal{R}$ was 4.13~pm/V with a standard deviation $\sigma_\mathcal{R}=0.29$~pm/V (see also Fig.~\ref{fig:tempdep}(c)).
The errors are discussed in detail in Section~\ref{sec:efsc}.
%The found value of $\mathcal{R}$ 
As is evident from Eq.~\ref{eq:E}, the constant value of the Pockels coefficient $\mathcal{R}$ clearly indicates a direct ratio of the difference between the phase shift difference $\Delta\Gamma_\mathrm{Pockels}(x_3)$ and the amplitude of the electric field $\mathcal{E}(x_3)$.

%Nasleduje pasaz o vlivu laseru

\subsubsection{Influence of the laser illumination and phase correction}

\begin{figure}[b!]%subfigure
    \begin{minipage}{.42\textwidth}
        \includegraphics[width=1\linewidth]{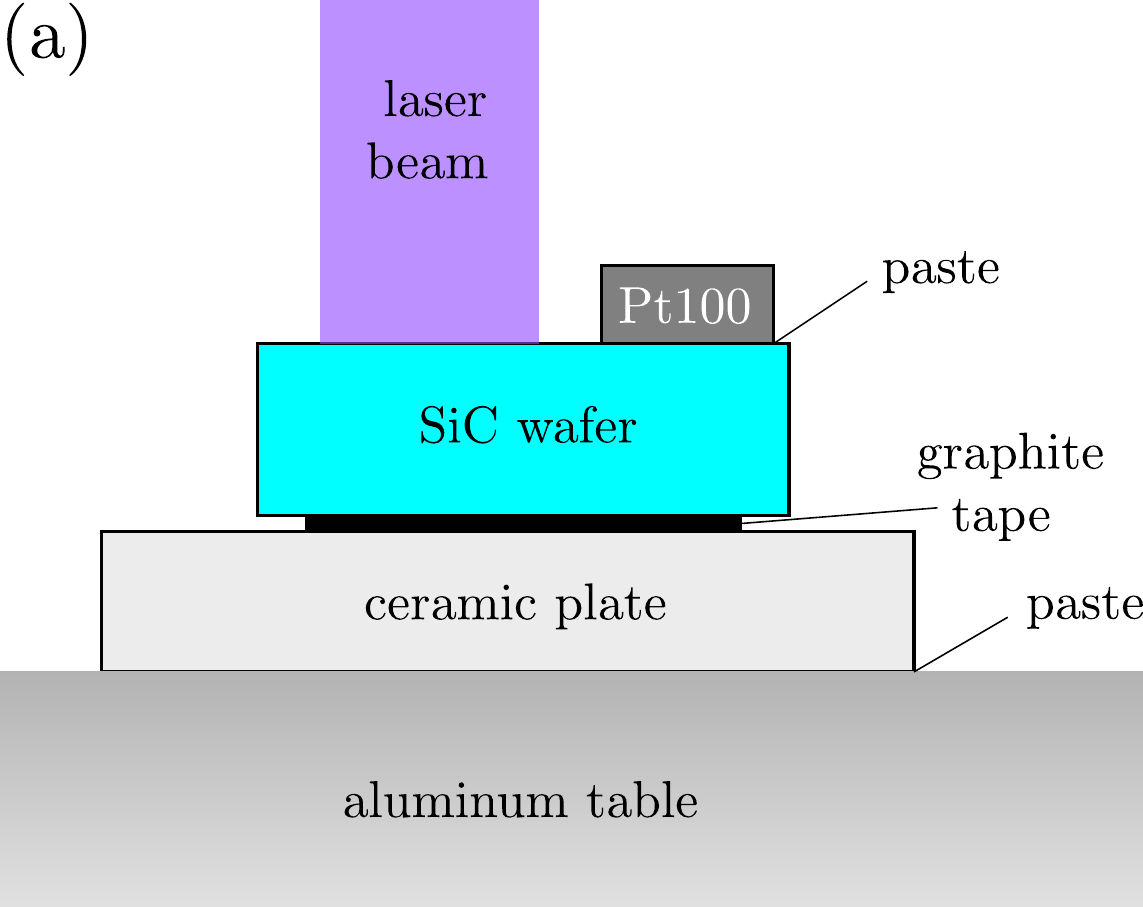}
    \end{minipage}
    \hfill
    \begin{minipage}{.53\textwidth}
        \includegraphics[width=1\linewidth]{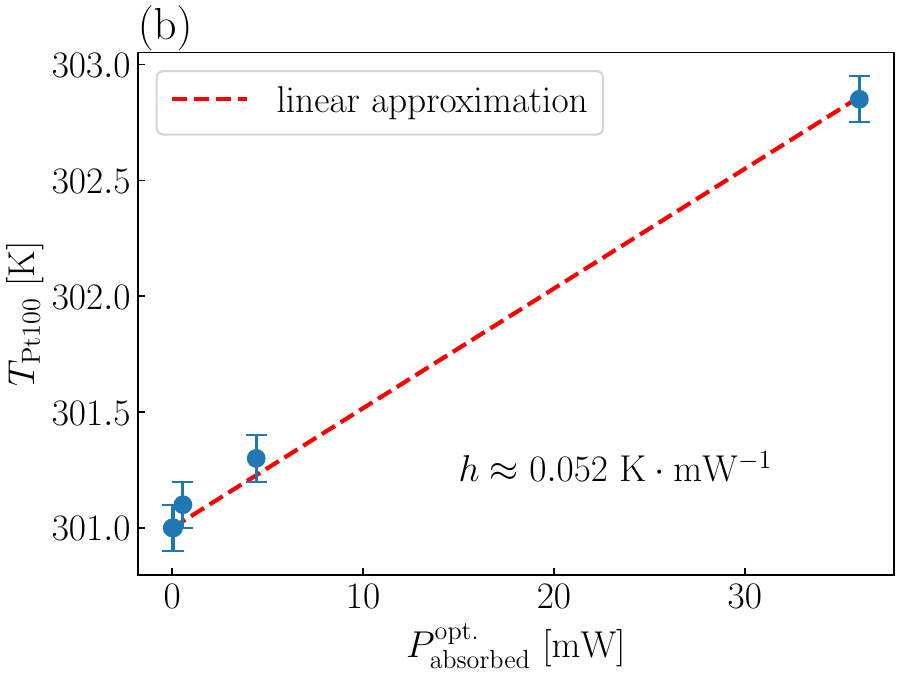}
    \end{minipage}
\caption{Scheme of the experimental setup used to measure the increase in the temperature in SiC under 405~nm laser illumination (a). Thermal paste was used to fix the Pt100 temperature sensor to the SiC wafer and the ceramic plate to the aluminum table. (b) shows the linear approximation (with slope $h$) of the temperature at Pt100 dependency on the absorbed laser power.}\label{fig:termo_setup}
\end{figure}

To investigate the influence of 405~nm laser illumination on the phase shift distribution difference $\Delta\Gamma_\mathrm{thermo-opt.}(x_3)$ caused by warming of the sample, a standard platinum resistance thermometer Pt100 was mounted on the top of the second 6H-SiC sample without electrodes (see Fig.\ref{fig:termo_setup}(a)). The sample was fixed to the ceramic plate with a graphite tape.
The ceramic plate was placed on an aluminum table, which ensured sufficient heat dissipation. For practical reasons and to prevent damage to the graphene and wire contact, this experiment was not performed on the first sample in the cryostat. We note that for these reasons the two samples differ, e.g. by the thermal contact to the ceramic plate.

The sample was illuminated by the 405~nm laser with power densities $S_1$ to $S_4$ and the temperature $T_{\mathrm{Pt100}}$ of the Pt100 sensor was monitored. It turned out that under the higher illumination, the sample started to heat up and, due to the temperature stabilization, its temperature stabilizes approximately within one minute. The observed temperature dependence on the absorbed optical power is shown in Fig.~\ref{fig:termo_setup}(b) and in Table~\ref{tab:tabulka}(a). Under the strongest illumination (power density $S_4$), the temperature of the sample $T_{\mathrm{Pt100}}$ increases by almost two kelvins. 

\begin{figure}[ht!]
\centering\includegraphics[width=1\textwidth]{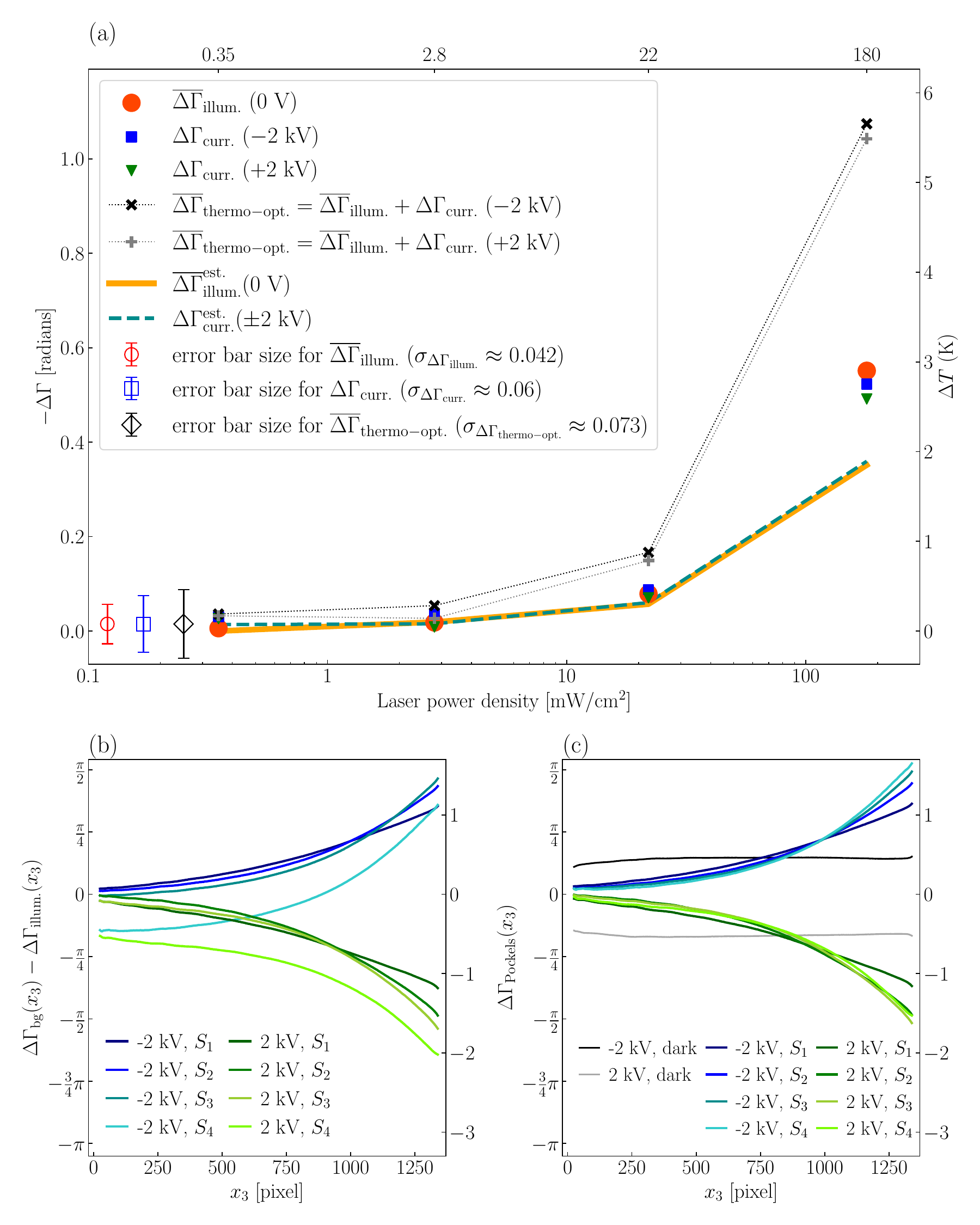}
\caption{Comparison of phase shift differences subtracted from phase shift measurements (separate points) and those estimated based on the temperature measurements (node points of thick curves) (a). Right axis shows the corresponding temperature increase $\Delta T=\Delta\Gamma /a$. Phase shift difference profiles cleaned by data measured under illumination at zero bias (b). Phase shift difference profiles related to the Pockels effect (c).}
\label{bigfig2}
\end{figure}

The corresponding estimated phase shift differences $\overline{\Delta\Gamma}_{\mathrm{illum.}}^\mathrm{est.}$ given by the thermo-optical phase shift coefficient $a=-0.19~\mathrm{K}^{-1}$ as $\overline{\Delta\Gamma}^\mathrm{est.}_\mathrm{illum.}=a(T_\mathrm{Pt100}-301$~K) are shown in Fig.~\ref{bigfig2}(a) as node points of the orange line and in Table~\ref{tab:tabulka}(a). Averaged experimental values (through the $x_3$-direction) $\overline{\Delta\Gamma}_\mathrm{illum.}$ of phase-shift difference profiles ${\Delta\Gamma}_\mathrm{illum.}(x_3)$ at zero bias and under the laser illuminations $S_1$ to $S_4$ from Fig.~\ref{bigfig1}(b) are shown as red circles in Fig.~\ref{bigfig2}(a). A relatively good agreement is apparent for both within 30\% error. Moreover, if we take into account that both experiments were performed under slightly different conditions, we can clearly attribute the light phase change under stronger laser illumination to the heating of the sample, which is almost homogeneous within the sample volume (see profiles (0~V, $S_3$) and (0~V, $S_4$) in Fig.~\ref{bigfig1}(b)).  As the value of $\overline{\Delta\Gamma}_\mathrm{illum.}$ and the $\Delta\Gamma_\mathrm{illum.}(x_3)$ profiles are calculated in the same way as the background profiles $\Delta\Gamma_\mathrm{bg}(x_3)$, their experimental error $\sigma_{\Delta\Gamma_\mathrm{illum.}}$ equals $\sigma_{\Delta\Gamma_\mathrm{bg}}\approx0.042$~radians.

\begin{table}
\caption{Temperature variation of $\Delta\Gamma$ in illuminated SiC}
\begin{center}
\begin{tabular}{ccccc}
\multicolumn{5}{l}{(a) Influence of illumination (unbiased sample)}\\						
\hline\hline							
&$S$ [mW/cm$^2$]	&	$P^{\mathrm{opt.}}_{\mathrm{absorbed}}$ [mW]	&	$T_\mathrm{Pt100}$ [K]	&	$\overline{\Delta\Gamma}_{\mathrm{illum.}}^\mathrm{est.}$ (0~V)$^\ast$	\\
\hline							
dark    & 0	&	0	&	301.0(1)	&	0	\\
$S_1$   &0.35	&	0.07	&	301.0(1)	&	0	\\
$S_2$   &2.8	&	0.56	&	301.1(1)	&	-0.019	\\
$S_3$   &22	&	4.4	&	301.3(1)	&	-0.057	\\
$S_4$   &180	&	36	&	302.9(1)	&	-0.352	\\
\hline	\hline
\multicolumn{5}{l}{$^\ast) \overline{\Delta\Gamma}^\mathrm{est.}_\mathrm{illum.}=a(T_\mathrm{Pt100}-301$~K)}\\
&&&	\\
\end{tabular}\\
\begin{tabular}{cccccc}						
\multicolumn{6}{l}{(b) Influence of the electric current (sample bias $V=-2$~kV)}\\						
\hline\hline								
&$S$ [mW/cm$^2$]	&	$I$ [\textmu A]	&	$P^\mathrm{el.}=V\cdot I$ [mW]	&$\Delta T^\mathrm{est.}$ [K]$^\dagger$	& $\Delta\Gamma_{\mathrm{curr.}}^\mathrm{est.}$ (-2~kV)$^\ast$	\\
\hline							
dark    &0	&	-0.7	&	1.4	&	0.073	&	-0.014	\\
$S_1$   &0.35	&	-0.775	&	1.55	&	0.081	&	-0.015	\\
$S_2$   &2.8	&	-1.04	&	2.08	&	0.11	&	-0.021	\\
$S_3$   &22	&	-3.05	&	6.1	&	0.32	&	-0.06	\\
$S_4$   &180	&	-18.2	&	36.4	&	1.89	&	-0.36	\\
\hline\hline
\multicolumn{6}{l}{$^\dagger$) $\Delta T^\mathrm{est.}=hP^\mathrm{el.}$,~~$^\ast$) $\Delta\Gamma^\mathrm{est.}_\mathrm{curr.}=a\Delta T^\mathrm{est.}$ }\\
\end{tabular}
\end{center}
\label{tab:tabulka}
\end{table}

Profiles ${\Delta\Gamma}_\mathrm{illum.}(x_3)$ corresponding to the respective laser power at zero bias can then be subtracted from the corresponding profiles $\Delta\Gamma_\mathrm{bg}(x_3
)$ measured under nonzero bias.  Subtracted phase shift difference profiles are shown in Fig.~\ref{bigfig2}(b) from which it is apparent that profiles except those at the highest laser power $S_4$ fulfill the condition of the constant integral (see Eq.~\ref{eq:calib}). In the next paragraphs, we focus on the explanation of the remaining shift in phaseshift difference in the case of the high laser power, which we attribute to the thermal power $P^\mathrm{el.}$ of the electric current.

Table~\ref{tab:tabulka}(b) shows the electric current $I$ flowing through the biased sample with graphene electrodes. Here, we focus only on the bias $V=-2$~kV, because for $+2$~kV we got almost symmetric values. Considering that the absorbed optical power $P^{\mathrm{opt.}}_{\mathrm{absorbed}}$ is completely converted into heat, we can obtain a power-temperature coefficient $h\approx 0.052\;\mathrm{K\cdot mW}^{-1}$ from the slope of the linear approximation of data points from the Fig.~\ref{fig:termo_setup}(b). The coefficient $h$ describes the increase in sample temperature with the absorbed optical power. We apply the same coefficient to estimate the increase in temperature $\Delta T^\mathrm{est.}$ due to the thermal power of the electric current shown in Table~\ref{tab:tabulka}(b). The estimated electric-current-induced difference of phase shift $\Delta\Gamma_{\mathrm{curr.}}^\mathrm{est.}=a\Delta T^\mathrm{est.}$ is shown in Table~\ref{tab:tabulka}(b) and depicted as node points of the green dashed line in Fig.~\ref{bigfig2}(a).

Independently of the estimation described above of $\Delta\Gamma_{\mathrm{curr.}}^\mathrm{est.}$, we subtracted the constant values of the difference $\Delta\Gamma_\mathrm{curr.}$ of the phase shift from the experimentally obtained profiles $\Delta\Gamma_\mathrm{bg}(x_3)-\Delta\Gamma_\mathrm{illum.}(x_3)$ from Fig.~\ref{bigfig2}(b) to meet the calibration condition from Eq.~\ref{eq:calib}. Values $\Delta\Gamma_\mathrm{curr.}$ are shown as blue squares and green triangles for both polarities in Fig.~\ref{bigfig2}(a). Relatively good match with the estimated values $\Delta\Gamma_{\mathrm{curr.}}^\mathrm{est.}$ based on the temperature and current measurements is apparent. The correlation in the magnitudes of phaseshift differences related to illumination and to the electric current is purely incidental, and both magnitudes are added to the thermo-optical contribution. As the value of $\Delta\Gamma_\mathrm{curr.}$ originates in subtracting $\Delta\Gamma_\mathrm{bg}(x_3)-\Delta\Gamma_\mathrm{illum.}(x_3)$ from the constant value, its error $\sigma_{\Delta\Gamma_\mathrm{curr.}}$ is given by $\sigma_{\Delta\Gamma_\mathrm{curr.}}=\sqrt{\sigma^2_{\Delta\Gamma_\mathrm{bg}}+\sigma^2_{\Delta\Gamma_\mathrm{illum.}}}=\sqrt{2\sigma^2_{\Delta\Gamma_\mathrm{bg}}}\approx0.06$~radians.

The phaseshift difference profiles $\Delta\Gamma_\mathrm{Pockels}(x_3)$ that we exclusively attribute to the Pockels effect are shown in Fig.~\ref{bigfig2}(c).
Here we note that it is impossible to determine the spatial $x_3$-distribution of $\Delta\Gamma_\mathrm{curr.}$, and therefore only a constant value can be considered. 
Then the total thermo-optic term $\Delta\Gamma_\mathrm{thermo-opt.}(x_3)$ is given by 
\begin{equation}
\Delta\Gamma_\mathrm{thermo-opt.}(x_3)\approx\Delta\Gamma_\mathrm{illum.}(x_3)+\Delta\Gamma_\mathrm{curr.},\label{eq:to}
\end{equation}
in which $\Delta\Gamma_\mathrm{illum.}(x_3)$ corresponds purely to the absorbed optical power, while $\Delta\Gamma_\mathrm{curr.}$ is related to the thermal power of the electric current which originates in the increase of sample's photoconductivity. This drop in the resistance of the sample under stronger illumination causes an increase in the electric current supplied by the sourcemeter to maintain the set voltage of $\pm2$~kV.
Averaged experimental values $\overline{\Delta\Gamma}_\mathrm{thermo-opt.}$ are also shown in Fig.~\ref{bigfig2}(a) as dark crosses for both polarities as well as its error $\sigma_{\Delta\Gamma_\mathrm{thermo-opt.}}=\sqrt{\sigma^2_{\Delta\Gamma_\mathrm{illum.}}+\sigma^2_{\Delta\Gamma_\mathrm{curr.}}}=\sqrt{3\sigma^2_{\Delta\Gamma_\mathrm{bg}}}\approx0.073$~radians.

Combining Eqs.~\ref{eq:gamy} and~\ref{eq:to}, the Pockels phase component $\Delta\Gamma_\mathrm{Pockels}(x_3)$ is then
\begin{equation}
\Delta\Gamma_\mathrm{Pockels}(x_3)\approx\Delta\Gamma_\mathrm{bg}(x_3)-\Delta\Gamma_\mathrm{illum.}(x_3)-\Delta\Gamma_\mathrm{curr.},
\end{equation}
in which thermo-optic-related components have almost constant distributions and cause the constant horizontal shift of the phase shift difference distribution $\Delta\Gamma_\mathrm{bg}(x_3)$ that can be relatively easily subtracted based on the calibration integral from the Eq.~\ref{eq:calib}. In the case of low illumination intensity and low electric current, one or both thermo-optic phase shift difference components can be neglected. Then the evaluation of the error related to the Pockels effect $\sigma_{\Delta\Gamma_\mathrm{Pockels}}$ depends on the total number of used components as shown in Table~\ref{tab:error}.

\begin{table}
\caption{Evaluation of $\Delta\Gamma_\mathrm{Pockels}(x_3)$ errors (sample bias $\pm2$~kV)}
\begin{center}
\begin{tabular}{ccc}	
\hline\hline
Conditions & \begin{tabular}{c}Components used for\\ $\Delta\Gamma_\mathrm{Pockels}(x_3)$ evaluation\end{tabular} & $\sigma_{\Delta\Gamma_\mathrm{Pockels}}$~[radians]\\
\hline\renewcommand{\arraystretch}{1.8}
dark &$\Delta\Gamma_\mathrm{bg}(x_3)$&$\sigma_{\Delta\Gamma_\mathrm{bg}}\approx0.042$\\
$S_1$, $S_2$ &$\Delta\Gamma_\mathrm{bg}(x_3)$, $\Delta\Gamma_\mathrm{illum.}(x_3)$ &$\sqrt{2\sigma^2_{\Delta\Gamma_\mathrm{bg}}}\approx0.06$\\
$S_3$, $S_4$ &$\Delta\Gamma_\mathrm{bg}(x_3)$, $\Delta\Gamma_\mathrm{illum.}(x_3)$, $\Delta\Gamma_\mathrm{curr.}$&$\sqrt{3\sigma^2_{\Delta\Gamma_\mathrm{bg}}}\approx0.073$\\
\hline\hline
\label{tab:error}
\end{tabular}
\end{center}
\end{table}

Since it is possible to satisfactorily explain the observed changes in the refractive index with the help of Pockels and the thermo-optical effects, photorefraction and the change of the refractive index with the free carrier concentration~\cite{Bennett1990, Bulutay2010}, were not taken into account because they are related to much higher laser powers and for materials with higher electrical conductivity, respectively.

\begin{comment}
\textbf{Influence of photorefraction:} To estimate the influence of the photorefraction through the Pockels effect, we can use the maximum power density $S_4$ as the mean value of the Poyinting vector magnitude of the laser light electromagnetic wave. Then, if we consider the wave polarization perpendicular to the $c-$axis, the amplitude $E_0$ of the electric field of this wave is
\begin{equation}
    E_0=\sqrt{\dfrac{2S_4}{n_o^\mathrm{405}c\varepsilon_0}},
\end{equation}
in which $n_o^\mathrm{405}\approx2.76$ is the ordinary refractive index of 6H-SiC at 405~nm~\cite{Wang2013}, $c$ is the vacuum speed of light and $\varepsilon_0$ vacuum permittivity. This yields $E_0\approx7\times10^{-4}~\mathrm{kV\cdot mm^{-1}}$. Although the geometry in this case is slightly different and the derivation of the suitable index ellipsoid for the electric field in the direction perpendicular for the $c-$axis would be necessary to exactly describe the effect of the photorefraction, its effect will be safely negligible because normally we observe electric field amplitude $\mathcal{E}$ ranging between approximately $0.1$ and $7~\mathrm{kV\cdot mm^{-1}}$ with error of $0.05~\mathrm{kV\cdot mm^{-1}}$ (see later results shown in Fig.~\ref{bigfig3}(a)).
\end{comment}

\subsection{Electric field and space charge evaluation}\label{sec:efsc}
\begin{figure}[ht!]
\centering\includegraphics[width=1\textwidth]{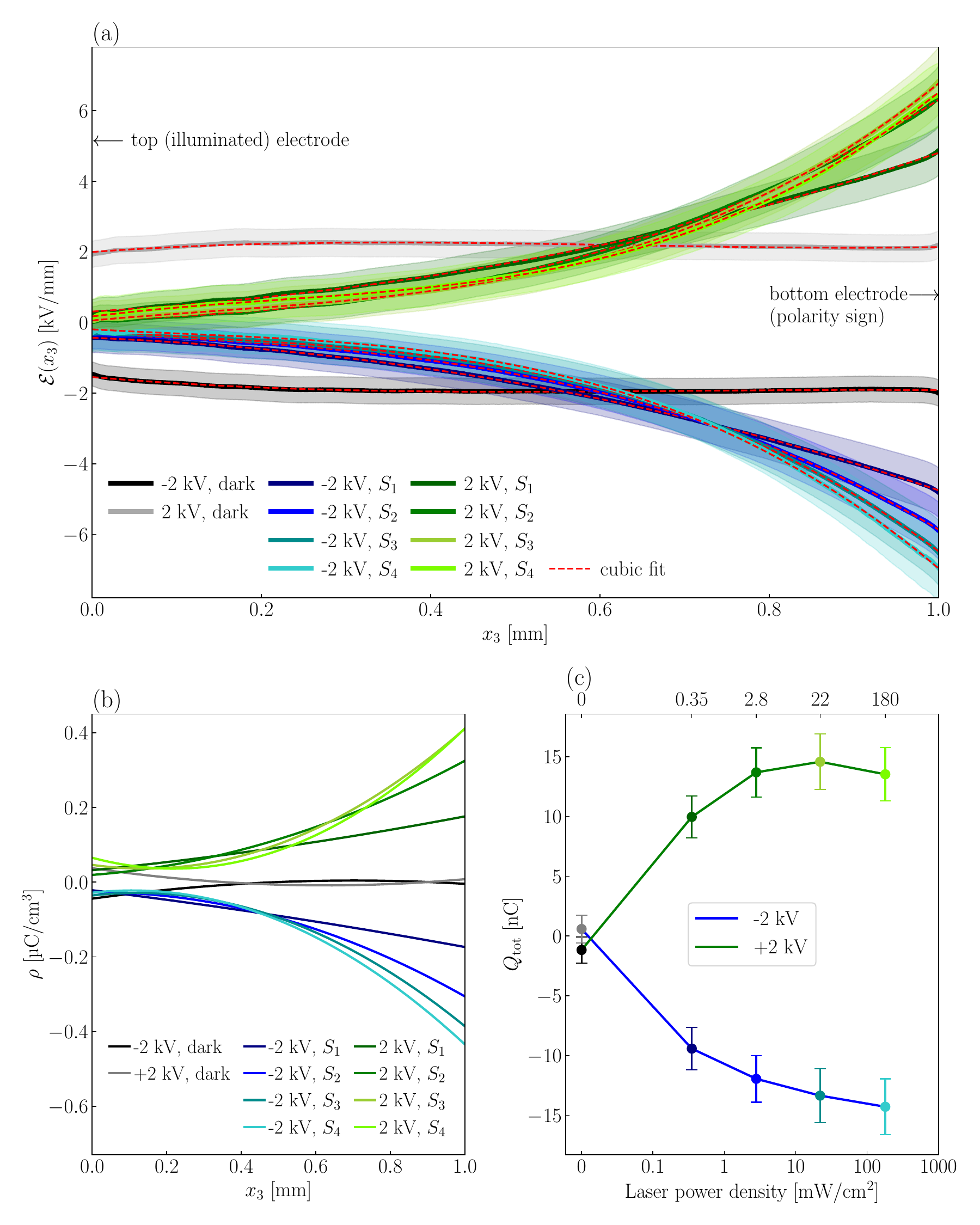}
\caption{The resulting electric field profiles (solid curves) with indicated standard deviations (thick semi-transparent bands) and their approximations by cubic function (red dashed curves) (a). Profiles of the space-charge density calculated from cubic fits (b). The bias polarity is related to the electrode at position $x_3=1$~mm. Evolution of the total space charge with laser power density (c).}
\label{bigfig3}
\end{figure}

The internal electric field profiles $\mathcal{E}(x_3)$ are calculated based on the Eq.~\ref{eq:E} from profiles  $\Delta\Gamma_\mathrm{Pockels}(x_3)$ (shown in Fig.~\ref{bigfig2}(c)) with known coefficient $\mathcal{R}=4.13$~pm/V found earlier (Section~\ref{sec:R}). Fig.~\ref{bigfig3}(a) shows that the $\mathcal{E}(x_3)$-profiles at zero bias without illumination are almost constant. The profiles $\mathcal{E}(x_3)$ under 405~nm laser illumination started to bent. While the electric field decreases under the illuminated electrode ($x_3=0$~mm), it increases under the bottom electrode ($x_3=1$~mm). This effect increases with the illumination intensity. Such behavior of electric field profiles in semi-insulating crystals under optical or X-ray irradiation is not surprising and is related to the space charge as recently reported by several research groups focused on radiation detectors made of Cd$_{1-x}$Zn$_x$Te, $x\in\langle0,0.12\rangle$,  crystals with $\overline{4}3m$ symmetry~\cite{Sellin2010, Prekas2010,Washington2011,Cola2013,Zazvorka2014,Cola2014,Franc2015}.

To estimate the error of electric field profiles $\sigma_\mathcal{E}$, we consider Eq.~\ref{eq:E} and the errors of the phase shift difference $\sigma_{\Delta\Gamma_\mathrm{Pockels}}$ from Table~\ref{tab:error} and of the Pockels coefficient $\sigma_\mathcal{R}=0.29$~pm/V. Then using Gauss’ law of error propagation we get
\begin{equation}
    \sigma_\mathcal{E}=\dfrac{\lambda_0}{2\pi d}\sqrt{\dfrac{1}{\mathcal{R}^2} \sigma^2_{\Delta\Gamma_\mathrm{Pockels}}+\dfrac{\Delta\Gamma_\mathrm{Pockels}^2}{\mathcal{R}^4}\sigma^2_\mathcal{R}},
\end{equation}
which scales with $\Delta\Gamma_\mathrm{Pockels}$ amplitude. %Both contributions ($\sigma_{\Delta\Gamma_\mathrm{Pockels}}$ and $\sigma_\mathcal{R}$) to $\sigma_\mathcal{E}$ are almost equal. 
In Fig.~\ref{bigfig3}(a), $\sigma_\mathcal{E}$ is represented by colored bands. 

In principle, it is possible to decrease $\sigma_{\Delta\Gamma_\mathrm{Pockels}}$ by phase measurements with better precision, that is, by using a camera with cooling to suppress its thermal noise and to use a stabilized coherent light source. It is further worth noting that the dark current in the sample at -2~kV (Table~\ref{tab:tabulka}(b)) induces a small but significant temperature-dependent change in the phase difference ($\Delta\Gamma^\mathrm{est.}_\mathrm{curr.}=-0.014$) compared to $\sigma_{\Delta\Gamma_\mathrm{Pockels}}$, which could cause a rather large uncertainty of $\sigma_\mathcal{R}=0.29$ (see Fig.~\ref{fig:tempdep}
(c)). Here, measuring the voltage dependence of phase differences in a sample with a higher resistance and thus a lower electric current could help to determine the coefficient $\mathcal{R}$ more precisely.

To determine a space charge distribution $\rho(x_3)$ from the electric field profiles $\mathcal{E}(x_3)$ we used the Gauss's law 
\begin{equation}
\rho(x_3)=\epsilon\nabla\cdot\mathcal{E}(x_3)=\epsilon\dfrac{\partial \mathcal{E}(x_3)}{\partial x_3},
    \label{eq:gauss}
\end{equation}
in which $\epsilon$ is the permittivity of 6H-SiC. To find a relatively smooth derivative of the $\mathcal{E}(x_3)$ profiles from Fig.~\ref{bigfig3}(a), it is necessary to smooth the data. It turned out that all these profiles can be well approximated by the cubic function, as shown by the red dashed curves in Fig.~\ref{bigfig3}(a). The resulting charge density profiles are shown in Fig.~\ref{bigfig3}(b). Total space charge $Q_\mathrm{tot}$ can be calculated from electric fields $\mathcal{E}_\mathrm{C}$ and $\mathcal{E}_\mathrm{A}$ at cathode and anode, respectively, as~\cite{Dedic2017} 
\begin{equation}
    Q_\mathrm{tot}=\epsilon A\left(\mathcal{E}_\mathrm{C}-\mathcal{E}_\mathrm{A}\right),
\end{equation}
 in which $A=0.25~\mathrm{cm}^2$ is the electrode area.
Evolution of $Q_\mathrm{tot}$ with laser 405~nm power at both polarities is shown in Fig.~\ref{bigfig3}(c). 
It is apparent that while for the positive polarity the total charge is small and negative in the dark condition and with increasing intensity of illumination it changes to positive and then increases, the opposite is true for the negative polarity. These phenomena are related to the trapping of photogenerated carriers at deep levels related to point defects of the crystal. However, since this paper primarily focuses on the method of the electric field measurement, a detailed explanation is out of scope and may be the subject of further studies.

\subsection{Method generalization}

The presented method will work on all identically oriented electric fields and crystals showing $6mm$ symmetry, in addition to 6H-SiC, i.e., on 4H-SiC and hexagonal polytypes of CdS, ZnS, and ZnO~\cite{Narasimhamurty1981}.
However, if we look at Eq.~\ref{eq:E}, thanks to the calibration condition (Eq.~\ref{eq:calib}) there is no need to know the value of the combination of electro-optical coefficients $\mathcal{R}$ in order to determine the electric field profiles in a crystal with the shape of a rectangular prism equipped with large planar electrodes on opposite sites. 
The method can therefore be applied to any Pockels material with a general crystal orientation except a few special cases due to inappropriate mutual orientation of the electric field, crystal, and testing light beam, when $\Delta\Gamma_\mathrm{Pockels}=0$ remains unchanged even if a voltage is applied to the sample. Such a situation is clearly explained, for example, in~\cite{Namba1961} for a ZnS crystal with symmetry $\overline{4}3m$. It is also necessary to keep in mind that different materials have different strengths of the Pockels effect (here, SiC belongs to the lowest average).
If the crystal orientation of the sample is unknown, it is appropriate to choose the rotation of the crossed polarizers so that the phase modulation caused by the shift of the SB compensator is the largest. Then it is guaranteed that the direction of easy passage of the polarizer will form an angle of 45 degrees between the axes of the elliptical section of the index ellipsoid~\cite{Dedic2021}.

\section{Conclusions}

The proposed method based on the Pockels effect allows to determine the internal electric field profiles in the biased and illuminated rectangular semi-insulating 6H-SiC crystal with planar electrodes on opposite sites based on the changes in mutual phase shifts of the polarized light. In addition to the Pockels effect, the heating of the crystal as a result of strong illumination has a fundamental effect on the phase shift and we distinguish these phenomena from each other. We estimated the contribution to the phase shift related to the Pockels effect by subtraction of a constant phase to meet the calibration of the resulting electric field profile integral to the bias applied to the crystal. 
The space charge distribution in the crystal can be calculated from electric fields. The studied SiC sample with graphene electrodes showed an increase in space charge under 405~nm laser illumination (up to 15~nC of total space charge under an optical power density of 180~mW/cm$^2$ in the studied sample with volume of 25~mm$^3$). This result brings new insight into the behavior of SiC in the presence of high-intensity light and has potential for further research and applications in optoelectronics.
The method can be adapted to any Pockels material with suitable geometry. It also allows finding the Pockels coefficients (or their combination, namely $\frac{1}{2}(n_e^3r_{33}-n_o^3r_{13})=4.13(29)$~pm/V at the wavelength of 546.1~nm for 6H-SiC) and gives insight into the study of the thermo-optical properties of the material.

\section{Backmatter}

\begin{comment}
Backmatter sections should be listed in the order Funding/Acknowledgment/Disclosures/Data Availability Statement/Supplemental Document section. An example of backmatter with each of these sections included is shown below.

\bmsection{Funding}
(Will be generated automaticaly)
\end{comment}
\begin{backmatter}
\bmsection{Acknowledgments}

\noindent The study was funded by the Czech Science Foundation (GAČR), project No.~22-20020S. We also acknowledge the CzechNanoLab Research Infrastructure supported by MEYS CR (LM2023051).

\bmsection{Disclosures}

\noindent The authors declare no conflicts of interest.

\bmsection{Data availability} Data underlying the results presented in this paper are not publicly available at this time due to their large volume ($\sim$10 GB of hundreds of high-resolution camera images) but may be obtained from the authors upon request.

\end{backmatter}

%%%%%%%%%% If using BibTeX:
\bibliography{sample}

%%%%%%%%%% If preparing manually:
% \begin{thebibliography}{1}
% \newcommand{\enquote}[1]{``#1''}

% \bibitem{Zhang:14}
% Y.~Zhang, S.~Qiao, L.~Sun, Q.~W. Shi, W.~Huang, L.~Li, and Z.~Yang,
%   \enquote{Photoinduced active terahertz metamaterials with nanostructured
%   vanadium dioxide film deposited by sol-gel method,}
%   {\protect\JournalTitle{Optics Express}} \textbf{22}, 11070--11078 (2014).

% \bibitem{Optica}
% {Optica}, \enquote{{Optica Publishing Group},}
%   \url{http://www.opg.optica.org}.

% \bibitem{FORSTER2007}
% P.~Forster, V.~Ramaswamy, P.~Artaxo, T.~Bernsten, R.~Betts, D.~Fahey,
%   J.~Haywood, J.~Lean, D.~Lowe, G.~Myhre, J.~Nganga, R.~Prinn, G.~Raga,
%   M.~Schulz, and R.~V. Dorland, \enquote{Changes in atmospheric consituents and
%   in radiative forcing,} in \enquote{Climate Change 2007: The Physical Science
%   Basis. Contribution of Working Group 1 to the Fourth assesment report of
%   Intergovernmental Panel on Climate Change,}  S.~Solomon, D.~Qin, M.~Manning,
%   Z.~Chen, M.~Marquis, K.~B. Averyt, M.~Tignor, and H.~L. Miler, eds.
%   (Cambridge University Press, 2007).

% \end{thebibliography}

\end{document}